\documentclass[preprint,showpacs,preprintnumbers,amsmath,amssymb]{revtex4}

\usepackage{graphicx}
\usepackage{dcolumn}
\usepackage{bm}

\begin{document}


\title{Conflict between the identification of cosmic neutrino source
and \\ the sensitivity to mixing angles in neutrino telescope}

\author{Ggyoung-Riun Hwang} \email{galaraja@phys.cau.ac.kr}
\author{Kim Siyeon} \email{siyeon@cau.ac.kr}

\affiliation{Department of Physics,
        Chung-Ang University, Seoul 156-756, Korea}
\begin{abstract}
Neutrino fluxes at telescopes depend on both initial fluxes out of
astronomical bursts and flavor mixing during their travel to the
earth. However, since the information on the initial composition
requires better precision in mixing angles and vice versa, the
neutrino detection at telescopes for itself cannot provide
solutions to the both problems. Thus, a probability to be measured
at long baseline oscillation is considered as a complement to the
telescope, and problems like source identification and parameter
degeneracy are examined under a few assumptions.
\end{abstract}


\maketitle \thispagestyle{empty}


\section{\label{sec:sec1}Introduction}

Cosmic neutrinos can be classified into stellar, galactic, and
extra-galactic neutrinos according to astronomical source. Stellar
neutrinos include solar and supernova neutrinos, of which energy
scale is order of (1-10)MeV since they are mainly produced by
nuclear interaction. There are fair records in which they were
identified as from the sun and SN1987A and examined for oscillation
and matter effect \cite{Ahmad:2002jz}\cite{Hirata:1987hu}. Ultrahigh
energy (UHE; $\gtrsim 10^{18}$eV) cosmic particles are regarded to
have their origin in extra-galactic source like active galactic
nuclei(AGN) and some of gamma ray bursts(GRB). Mechanisms to
accelerate protons to high energy have been searched, from
$\mathcal{O}$(PeV) to GZK limit, in GRB
\cite{Wick:2003ex}\cite{Waxman:1997ti}. In astronomy, GRB's are
detected once a day on average. If an accelerated proton produces a
pion, about 20$\%$ of its energy is transferred to neutrinos,
although more than one pion can be produced from a proton. Thus
there are abundant source of neutrinos whose energy is higher than
100TeV \cite{Rachen:1998fd}.

Decays of pions, such as $\pi^+ \rightarrow \mu^+ + \nu_\mu
\rightarrow e^+ + \nu_\mu + \nu_e + \bar{\nu}_\mu$ or $\pi^-
\rightarrow \mu^- + \bar{\nu}_\mu \rightarrow e^- + \nu_\mu +
\bar{\nu}_e + \bar{\nu}_\mu$, are the main process of neutrino
production. The initial neutrino flavor ratio can be determined,
depending on the charges of pions and the energy loss rates of
muons in processes of pion decays. Whether pions are produced by
$pp$ or by $p\gamma$ collision, the charges of pions and the ratio
of $\nu_e/\bar{\nu}_e$ are different. Although both pions and
muons lose energy in environment that pions were borne, long-lived
muons are more probable to interact with environment before
decaying. The pions that decay into neutrinos without muon decays
are called ``muon-damped source". Such electromagnetic energy loss
of muons for GRB becomes significant when the energy of
$\gamma$-ray $E_\gamma$ is $\gtrsim 100$TeV and, thus, the energy
of neutrino $E_\nu$ produced from muon-damped source is $\gtrsim
1$TeV \cite{Kashti:2005qa}. The intensities of neutrinos in
$10^{14}\rm{eV}<E<10^{16}\rm{eV}$ and higher energy were studied.
Its model-independent upper bound was predicted
\cite{Waxman:1997ti}\cite{Waxman:1998yy}. If the decay mode of a
pion includes muon decay, that will be called simply  ``pion
source" in comparison with muon-damped source. The initial
neutrino composition
$\Phi^0(\nu_e):\Phi^0(\nu_\mu):\Phi^0(\nu_\tau)=1:2:0$ from the
pion source is gradually replaced by the composition
$\Phi^0(\nu_e):\Phi^0(\nu_\mu):\Phi^0(\nu_\tau)=0:1:0$ from the
muon-damped source when the energy increases passing about 1TeV
\cite{Kashti:2005qa}\cite{Beacom:2003nh}. When the main source of
neutrino production in atmosphere is pion decay, the shift of the
ratio $\Phi^0(\nu_\mu)/\Phi^0(\nu_e)$ from 2 to infinity according
to increasing energy \cite{Gaisser:1988ar}\cite{Honda:1990sx} was
tested in SK \cite{Fukuda:1994mc}\cite{Fukuda:1998ub}. The
transition in the flavor ratio can be naturally assumed to occur
at the region of GRB's, and the corresponding aspect was discussed
in Ref. \cite{Kashti:2005qa}. The relative flavor ratios of
detected fluxes vs. energy exhibit saturation and transition. The
saturation of $\Phi^t(\nu_\mu)/\Phi^t(\nu_e)$ close to one implies
that the beam originated from pion source for lower energy, while
the transition of $\Phi^t(\nu_\mu)/\Phi^t(\nu_e)$ to a larger one
implies that the beam includes portion of muon-damped source for
higher energy.

There are a few neutrino telescopes under construction, which can
detect high energy neutrinos of $E_\nu > 0.1$TeV, e.g., IceCube
\cite{Ahrens:2002dv}, Antares \cite{Montaruli:2003bc}, etc. They
are designed to detect VHE and UHE neutrinos and to distinguish a
flavor from others. With a future Cerenkov detector, $\nu_\mu$ is
easy to be detected due to the long tracks of muons, while
$\nu_\tau$ is distinguishable from $\nu_e$ by double-bang event
only near PeV range. At $\mathcal{O}$(PeV), the neutrino
telescopes will be able to observe both double showers of
$\nu_\tau$, one due to $\tau$ production and the other due to
$\tau$ decay, and $W^-$ resonant event at 6.3PeV that identifies
$\bar{\nu}_e$ \cite{Learned:1994wg}.

For cosmic neutrinos, the oscillation factor in a transition
probability is averaged out due to long distance of travel and
high frequency. Thus, the telescope experiment can be effective
measurements of mixing parameters, provided that the neutrino
flavor ratio at the source is known \cite{Athar:2000yw}
The current data of mixing
parameters are phrased by the ranges in the magnitude of PMNS
elements;
\begin{eqnarray}
    |U_{PMNS}| = \left(
        \begin{array}{ccc}
        0.79-0.86 & 0.50-0.61 & 0-0.20 \\
        0.25-0.53 & 0.47-0.73 & 0.56-0.79 \\
        0.21-0.51 & 0.42-0.69 & 0.61-0.83
        \end{array}\right)
    \label{pmnssize}
\end{eqnarray}
at $3\sigma$ level \cite{Yao:2006px}. The matrix implies the
following values of individual parameters; $\Delta
m^2_{21}=(7.1-8.9)\times 10^{-5}\mathrm{eV}^2, ~|\Delta
m^2_{31}|=(2.2-3.0)\times 10^{-3}\mathrm{eV}^2,
~\sin^2\theta_{12}=0.24-0.40, ~\sin^2\theta_{23}=0.34-0.68,$ and
$\sin^2\theta_{13}\leq0.040$, all at $3\sigma$ level
\cite{Maltoni:2004ei}. The type of the yet-undetermined among
neutrino parameters can be organized case by case: First, normal
hierarchy (NH) or inverse hierarchy (IH): the sign of $\Delta
m^2_{31} \equiv m^2_3-m^2_1$ is unknown while the global best-fit
is obtained as $|\Delta m^2_{31}|=2.6\times10^{-3}\mathrm{eV^2}$.
Second, the sign of $\theta_{23}-\pi/4$, i.e., whether the
majority of $\nu_\tau$ is the heavier($\nu_3$ for NH) or the
lighter($\nu_2$ for NH) is unknown. Third, a certain value of
oscillation probability has infinite number of candidate
combination of $\theta_{13}$ and $\delta_{CP}$
\cite{Barger:2001yr}\cite{Minakata:2002jv}. The ambiguity due to
the above degeneracies needs to be clearly distinguished from that
due to uncertainties.

In Sec. \ref{sec:sec2}, the probability of a neutrino oscillation
is reviewed, focused on how the degeneracies are generated and
which oscillation experiment each mixing angle is most sensitive
to. In Sec. \ref{sec:sec3}, examined are the sensitivities of
neutrino fluxes to neutrino mixing angles. In Sec. \ref{sec:sec4},
we discussed a few points including parameter degeneracies and
source identification, where the data of neutrino telescopes are
considered in company with the data of a long baseline (LBL)
oscillation under a few assumptions. Concluding remarks follow in
Sec. \ref{sec:sec5}.

\section{\label{sec:sec2}Review on degenerate probability of oscillation}
The probability of neutrino oscillation may be said to be
degenerate because different sets of parameters result in the same
value. For example, the probability of transition from
$\nu_\alpha$ to $\nu_\beta$ in two-neutrino oscillation with a
single mixing angle and a single mass-squared difference $\Delta
m^2$
    \begin{eqnarray}
    P_{\alpha\beta}= \delta_{\alpha\beta}-(2\delta_{\alpha\beta}-1)
    \sin^22\theta\sin^2(\frac{\Delta m^2 L}{4E}),
    \end{eqnarray}
is invariant under switching the sign of $\Delta m^2$ or changing
the angle $\theta$ with its complementarity angle $\pi/2-\theta$.
Since the two sets, $(\theta,\Delta m^2)$ and
$(\pi/2-\theta,-\Delta m^2)$, are not physically different, the
above probability is twofold degenerate simply due to $(\theta,
\Delta m^2)$ and $(\pi/2-\theta, \Delta m^2)$ or due to $(\theta,
\Delta m^2)$ and $(\theta, -\Delta m^2)$.

The oscillation probability extended to three neutrinos in vacuum
\begin{footnotesize}
    \begin{eqnarray}
    P_{\alpha\beta}= \delta_{\alpha\beta} &-&
    4\sum_{i=1}^2\sum_{j=i+1}^3
    Re[U_{\alpha i}U^*_{\beta i}U^*_{\alpha j}U_{\beta j}]
    \sin^2(\frac{\Delta m^2_{ji} L}{4E}) \nonumber \\
    &\pm& 2\sum_{i=1}^2\sum_{j=i+1}^3
    Im[U_{\alpha i}U^*_{\beta i}U^*_{\alpha j}U_{\beta j}]
    \sin(\frac{\Delta m^2_{ji} L}{2E})
    \label{3osc}
    \end{eqnarray}
\end{footnotesize}
is given in terms of $3 \times 3$ unitary transformation matrix
$U$ and three mass-squared differences $\Delta m^2_{ji} \equiv
m_j^2-m_i^2$, where each of $\nu_\alpha$ and $\nu_\beta$ may be
one of $\nu_e, ~\nu_\mu$ or $\nu_\tau$. In effect $\Delta m^2_{31}
\simeq \Delta m^2_{32}$ so that they will not be distinguished
hereafter. The PMNS matrix in standard parametrization is given by
    \begin{footnotesize}
    \begin{eqnarray}
        U_{\alpha i} \equiv  \label{ckm}
        \left(\begin{array}{ccc}
            c_{12}c_{13} & s_{12}c_{13} & s_{13}e^{-i\delta}\\
            -s_{12}c_{23}-c_{12}s_{23}s_{13}e^{i\delta} &
            c_{12}c_{23}-s_{12}s_{23}s_{13}e^{i\delta} &
            s_{23}c_{13} \\
            s_{12}s_{23}-c_{12}c_{23}s_{13}e^{i\delta} &
            -c_{12}s_{23}-s_{12}c_{23}s_{13}e^{i\delta} &
            c_{23}c_{13}\
        \end{array}\right),\nonumber
    \end{eqnarray}
    \end{footnotesize}
where $s_{ij}$ and $c_{ij}$ denote $\sin{\theta_{ij}}$ and
$\cos{\theta_{ij}}$ with the mixing angle $\theta_{ij}$ between
$i$-th and $j$-th generations, respectively, and $\delta$ denotes
a Dirac phase.

The degeneracy of probability could be stemmed from three mixing
angles and three mass-squared differences. However, since a number
of successful experiments found the values of some physical
parameters, the possible multiplicity of degeneracy at present is
at most eight. A twofold degeneracy in the eightfold degeneracy
has its origin in the ambiguity of a value of $\theta_{23}$ from
its complementary angle, $\pi/2-\theta_{23}$. The
$\nu_\mu$-disappearance probability which is most sensitive to
determine the value of $\theta_{23}$ is expressed by
\cite{Minakata:2002jv}
\begin{footnotesize}
    \begin{eqnarray}
    &&
    1-P(\nu_\mu \rightarrow \nu_\mu) =
    \sin^22\theta_{23}\cos^2\theta_{13}
    \sin^2(\frac{\Delta m^2_{31}L}{4E_\nu}) \nonumber \\
    &&
    -~\sin(\frac{\Delta m^2_{21}L}{4E_\nu})
    \sin(\frac{\Delta m^2_{31}L}{4E_\nu})\sin^22\theta_{23}
    \cdot \nonumber \\
    &&
    \cdot(\cos^2\theta_{13}\cos^2\theta_{12} -
    \sin\theta_{13}\sin^2\theta_{23}\sin2\theta_{12}\cos\delta)
    , \label{disappear}
    \end{eqnarray}
\end{footnotesize}
where the leading two terms are invariant under the exchange of
$\theta_{23}$ with its complementary angle. On the other hand, the
$\nu_e$-appearance probability with matter effect $A$,
\begin{footnotesize}
    \begin{eqnarray}
    && P(\nu_\mu (\overline{\nu}_\mu)\rightarrow
        \nu_e (\overline{\nu}_e)) =
    \sin^2\theta_{23}\sin^22\theta_{13}
    \frac{\sin^2((\Delta m^2_{31}\mp A)L/4E_\nu)}
        {(1\mp A/\Delta m^2_{31})^2}  \nonumber \\
    &&
    + ~\frac{\Delta m^2_{21}}{\Delta m^2_{31}}
    \sin2\theta_{23}\sin2\theta_{13}\sin2\theta_{12}
    \cos(\delta\pm\Delta m^2_{31}L/4E_\nu)\cdot \nonumber \\
    &&
    \hspace{30pt}\cdot
    \frac{\sin((\Delta m^2_{31}\mp A)L/4E)\sin( A L/4E_\nu)}
    {(1\mp A/\Delta m^2_{31})( A/\Delta m^2_{31})}
    \nonumber \\
    &&
    + ~(\frac{\Delta m^2_{21}}{\Delta m^2_{31}})^2
    \cos^2\theta_{23}\sin^22\theta_{12}
    \frac{\sin^2( A/4E_\nu)}{( A/\Delta m^2_{31})^2}, \label{lblp}
    \end{eqnarray}
\end{footnotesize}
provides multi solutions to $(\theta_{13},\delta)$ pair and
$(\Delta m^2_{31},\delta)$ pair, where $A \equiv
2\sqrt{2}\mathrm{G_FY_e}\rho E_\nu$ is given with a density $\rho$
and an electron fraction $Y_e$
\cite{Barger:2001yr}\cite{Minakata:2002jv}. The leading term in
Eq. (\ref{lblp}) is not sensitive to the sign of $\Delta
m^2_{31}$, if the matter effect A is not significant. The second
term is also not indicative of the sign due to $\cos\delta$ and a
suppressing factor $\Delta m^2_{21}/\Delta m^2_{31}$. So the
$(\Delta m^2_{31},\delta)$ pair causes another double degeneracy
to the probability. It is well known as a matter of normal
hierarchy or inverse hierarchy. The multiple possibilities in the
pair of $(\theta_{13}, \delta)$ for a value of
$\sin2\theta_{13}\cos\delta$ make up the eight-fold degeneracy
with other two double degeneracies.

If astronomical neutrinos from origins like grb's or other types of
extragalactic bursts are considered for the flavor transition, in
the limit $L \rightarrow \infty$, the probability in Eq.
(\ref{3osc}) reduces to
    \begin{eqnarray}
    P^t_{\alpha\beta} &\rightarrow& \delta_{\alpha\beta}-
    2\sum_{i=1}^2\sum_{j=i+1}^3
    Re[U_{\alpha i}U^*_{\beta i}U^*_{\alpha j}U_{\beta j}] \\
    &=& \sum_{i=1}^3 |U_{\alpha i}|^2|U_{\beta i}|^2. \label{teleprob}
    \end{eqnarray}
Since the probability of relatively long-distance oscillation
compared to wavelength is averaged out, the dependency in the sign
of $\Delta m^2_{ji}$ is hidden, so that the probability is blind
to $\Delta m^2_{ji}$. Since it depends on only the absolute values
of the elements in PMNS as shown in Eq. (\ref{teleprob}), its
dependency on the phase $\delta$ is simply $\cos\delta$. Sensitive
dependencies between other mixing angles and $P^t_{\alpha\beta}$
will be discussed in detail in Section IV.

Degeneracy problems are caused since the number of effective
measurements is not sufficient to specify all the physical
parameters and the forms of probabilities are multi-variable
sinusoidal functions. Here, the ambiguities from three types of
degeneracies, the sign of $\Delta m^2_{31}$
\cite{Minakata:2001qm}, the sign of $\pi/4-\theta_{23}$
\cite{Fogli:1996pv}, and different pairs of $(\theta_{13},
\delta)$'s \cite{BurguetCastell:2001ez} will be clearly
distinguished from the ambiguities from uncertainties. The
degeneracy is a problem causing ambiguity even when an average
probability or an average flux without uncertainty is applied to
determine parameters. For more review, see references
\cite{GonzalezGarcia:2007ib}.

\section{\label{sec:sec3}Limit of neutrino telescope}

The initial flux of astronomical high energy neutrinos is assumed
to be attributed from the decay of pion produced in $p\gamma$
collision. The relative ratio of neutrino flavors from pion decay
is $\Phi^0(\nu_e):\Phi^0(\nu_\mu):\Phi^0(\nu_\tau)=1:2:0$. If the
daughter muon in pion decay does not decay due to the
electromagnetic energy loss, the flavor ratio is changed into
$\Phi^0(\nu_e):\Phi^0(\nu_\mu):\Phi^0(\nu_\tau)=0:1:0$. So, the
composition ratio $\Phi^0(\nu_\mu)/\Phi^0(\nu_e)=2$ and its
transition to infinity as the energy of neutrinos increases may
occur in region of GRB's, as does in atmosphere. However, it is
impossible to estimate the initial flux of neutrinos as being
produced in GRB's. One can but conjecture it from detected fluxes
at telescopes \cite{Barenboim:2003jm}. If the neutrino mixing is
of tri-bi maximal type as the simplest example, the pion source is
identified by the flux ratio at telescopes,
$\Phi^t(\nu_e):\Phi^t(\nu_\mu):\Phi^t(\nu_\tau)=1:1:1$, while the
muon-damped source is identified by the flux ratio at telescopes,
$1:1.8:1.8$. Thus, the transition in flavor ratios of detected
fluxes at GRB's occurs with the relative ratio
$\Phi^t(\nu_\mu)/\Phi^t(\nu_e)$ from 1 to 1.8 as the energy of
neutrinos increases. In reality, it is impossible to determine,
simply by reading results at telescopes, whether initial beams are
pion source or muon-damped source, without any assumption, for the
following reasons: First, any information from astronomical
neutrino bursts on the composition in a mixture of two types of
sources cannot be obtained without using telescopes. Second, the
mixing angles and masses are still bearing too broad uncertainties
to analyze the fluxes, and moreover the fluxes to be detected at
telescopes are significantly sensitive to mixing angles.
Hereafter, we took a strong assumption which is that pion source
and muon-damped source can be distinguished, in order to
illustrate the point of the sensitivity to mixing angles.

\begin{figure}
{\includegraphics[width=0.75\textwidth]{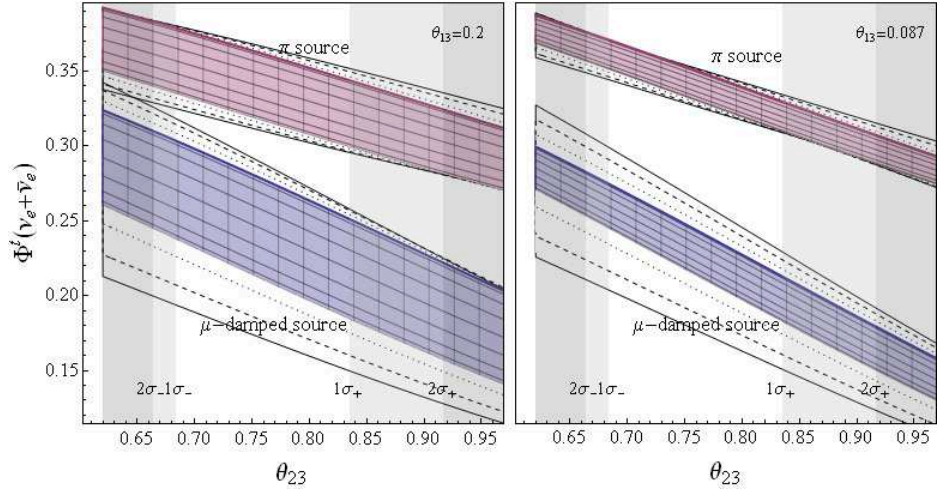}}
\caption{\label{fig:twosonu} $\Phi^t(\nu_e)$ vs. $\theta_{23}$ for
fixed values of $\theta_{13}$.}
\end{figure}

\begin{figure}
{\includegraphics[width=0.75\textwidth]{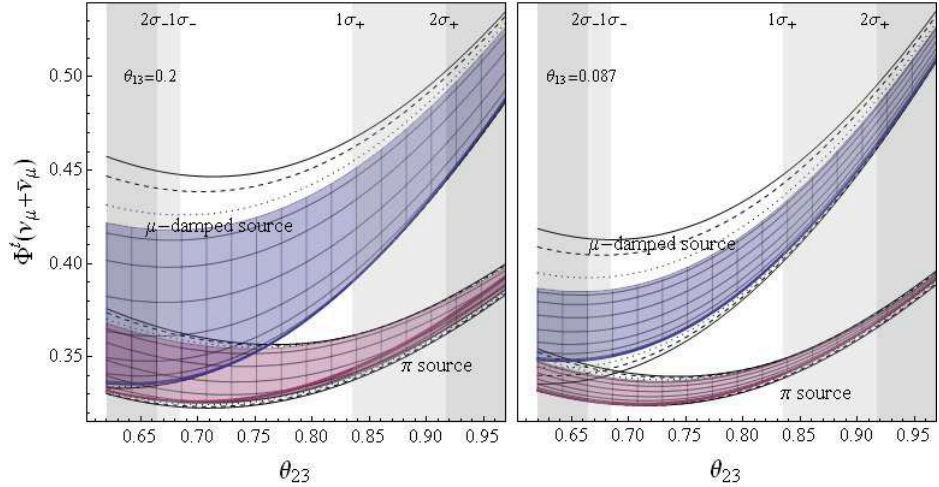}}
\caption{\label{fig:twosomu} $\Phi^t(\nu_\mu)$ vs. $\theta_{23}$ for
fixed values of $\theta_{13}$.}
\end{figure}

Unless neutrinos decay,
$\Phi^0(\nu_e)+\Phi^0(\nu_\mu)+\Phi^0(\nu_\tau)=
\Phi^t(\nu_e)+\Phi^t(\nu_\mu)+\Phi^t(\nu_\tau)$ and ~$\sum _\alpha
P^t_{\alpha\beta} =1$, where $\Phi^t(\nu_\beta)=\sum _\alpha
P^t_{\alpha\beta} \Phi^0(\nu_\alpha)$. Since relative fluxes at a
telescope can be normalized such that $\sum_\alpha
\Phi^0(\nu_\alpha)=\sum_\alpha \Phi^t(\nu_\alpha)=1$, the
normalized flux $\Phi^t(\nu_\alpha)$ is a linear combination of
$P^t_{\beta\alpha}$'s, for example,
$\Phi^t(\nu_\alpha)=P^t_{\mu\alpha}$ for muon-damped source and
$\Phi^t(\nu_\alpha)=1/3(P^t_{e\alpha}+2P^t_{\mu\alpha})$ for pion
source. The symmetric matrix $P^t_{\alpha\beta}$ in Eq.
(\ref{teleprob}) has only three independent elements. Thus, when
the initial condition of neutrino beam is assumed such that both
pure pion source and pure muon-damped source are allowed,
$\Phi^t(\nu_\tau)$ and one of the following four fluxes may be
unnecessary: $\Phi^t(\nu_e)$ from pion source, $\Phi^t(\nu_\mu)$
from pion source, $\Phi^t(\nu_e)$ from muon-damped source, and
$\Phi^t(\nu_\mu)$ from muon-damped source. In reality, either
$\Phi^t(\nu_e)$ or $\Phi^t(\nu_\mu)$ is more likely to be a mixed
flux from pion source and muon-damped source, rather than a flux
from the pure pion source or the pure muon-damped source. In order
to examine the sensitivities to mixing angles, however, we
consider the pure pion or the pure muon-damped source. FIG.
\ref{fig:twosonu} and FIG. \ref{fig:twosomu} illustrate the
dependence of $\Phi^t(\nu_e)$ on $\theta_{23}$ and the dependence
of $\Phi^t(\nu_\mu)$ on $\theta_{23}$, respectively. For a given
$\theta_{13}$, the range of $\delta$ from 0 to $2\pi$ makes the
curves thick bands. The red (upper in Fig. 1 and lower in
Fig.2)and blue (lower in Fig.2 and upper in Fig.1) bands indicate
the fluxes from the pure pion source or from the pure muon-damped
source, respectively.


The symmetric probability in Eq. (\ref{teleprob}) has ranges
estimated from PMNS at $3\sigma$ CL in Eq. (\ref{pmnssize});
\begin{eqnarray}
    P_{\alpha\beta} = \left(
        \begin{array}{ccc}
        0.48-0.64 & 0.12-0.34 & 0.11-0.35 \\
        \surd & 0.33-0.53 & 0.30-0.41 \\
        \surd & \surd & 0.33-0.47
        \end{array}\right).
    \label{prorange}
\end{eqnarray}
Thus, the fluxes accompany nontrivial ranges. The plots in FIG.
\ref{fig:twosonu} and FIG. \ref{fig:twosomu} illustrate the
sensitivity of fluxes to all the mixing angles including CP phase
$\delta$. The significant sensitivity to $\theta_{23}$ is expressed
by the broad coverage in $\Phi^t(\nu_e)$ and $\Phi^t(\nu_\mu)$ of
each colored stripe for fixed $\theta_{12}$ and $\theta_{13}$. Each
stripe is a bundle of curves for $\delta$ values from 0 to $\pi$.
The top line of the stripe for $\Phi^t(\nu_e)$ indicates $\delta=0$,
while the bottom line of the stripe for $\Phi^t(\nu_\mu)$ indicates
$\delta=0$ . Next, the sensitivity to $\theta_{12}$ was estimated
only for $\delta=0$ and $\delta=\pi$ and is illustrated by three
kinds of lines outside the colored. It is clear that the
$\theta_{12}$ does not affect the fluxes as much as the
$\theta_{23}$ does. The $\theta_{13}$ dependency in the fluxes is
illustrated by the widths of the flux stripes. A remarkable
improvement in precision of $\theta_{13}$ will be obtained from near
future oscillation experiments, e.g., Daya Bay \cite{Guo:2007ug},
Double CHOOZ \cite{Ardellier:2006mn}, RENO \cite{Joo:2007zzb},
No$\nu$A \cite{Ayres:2004js} and T2K \cite{Itow:2001ee}. In FIG.
\ref{fig:twosonu} and FIG. \ref{fig:twosomu}, two values of
$\theta_{13}$ are used for comparison: 0.20 which is the current
upper bound at $3\sigma$ CL and 0.082 which is the upper bound at
$3\sigma$ that RENO will accomplish. It is clearly illustrated that
narrowing the stripes by a small bound of $\theta_{13}$ makes
identifying the relative composition more realistic.

If a series of LBL oscillations, e.g., No$\nu$A and T2K, are
successful in achieving the aimed precision level, reducing the
relative range of $\sin^2\theta_{23}$ at $3\sigma$ from 79\% to
42\% \cite{Lindner:2005xf}, the range in $\Phi^t(\nu_e)$ from the
pure pion source is completely separated from the range in
$\Phi^t(\nu_e)$ from the pure muon-damped source. The estimation
of composition rate in a mixed beam of two pure sources requires
the determination of CP phase $\delta$ as well as improved
precisions of other parameters.

The curves of neutrino fluxes with respect to energy show both
saturation and transition. For instance, the initial flux ratio
$\Phi^0(\nu_\mu)/\Phi^0(\nu_e)$ of atmospheric neutrinos saturates
to 2 at low energy limit \cite{Gaisser:1988ar}\cite{Honda:1990sx},
while the expected ratio $\Phi^t(\nu_\mu)/\Phi^t(\nu_e)$ of high
energy cosmic neutrinos to be detected saturates to 1 at low
energy limit and to 1.8 at high energy limit \cite{Kashti:2005qa}.
If we can distinguish the saturation from the transition in flux -
energy plots, the identification of neutrino beam from pure pion
source or pure muon-damped source may be possible only after the
accumulation of sufficient data of cosmic neutrino detection.

\section{\label{sec:sec4}LBL: A complement to telescope}

The probability $P_{\mu e}$ in Eq. (\ref{lblp}) depends on
$\cos(\delta+\Delta m_{31}^2L/4E_\nu)$, while $\Phi^t(\nu)$ in Eq.
(\ref{teleprob}) depends on $\cos\delta$. If $\Phi^t(\nu)$ and
$P_{\mu e}$ are the orthogonal axes as in FIG. \ref{fig:eightpi},
the locus of $\Phi^t(\nu)$ and $P_{\mu e}$ for $\delta$ from 0 to
$2\pi$ completes a closed path. In FIG. \ref{fig:eightpi}, every
locus in each figure passes a point in $P_{\mu e}-\Phi^t(\nu_e)$
space. The phrases $P(\nu_\mu \rightarrow \nu_e)$ and
$\Phi(\nu_\alpha+\bar{\nu}_\alpha)$ in figures are equivalent to
$P_{\mu e}$ and $\Phi(\nu_\alpha)$ in text. Especially in these
figures, the probability $P_{\mu e}$ is an expected measurement of
a super beam from J-PARC to Super-Kamiokande (T2K collaboration).
Thus, input values are adopted from Ref. \cite{Itow:2001ee}. The
baseline is 295km and the energy of neutrino is 1GeV. The matter
effect $A$ in Eq. (5) is obtained with the Earth's density $ 2.8
\mathrm{g/cm^3}$ and the electron fraction $0.5$. The Eq.
(\ref{disappear}) and Eq. (\ref{lblp}) include only the leading
terms.

\subsection{Eightfold degeneracy}

A number of strategies are proposed to solve the degeneracy
problems by using the results of LBL oscillations and reactor
neutrino oscillations in future
\cite{Barger:2001yr}-\cite{GonzalezGarcia:2007ib}. An example is
to analyze long-baseline experiments over different oscillation
distances \cite{Barger:2001yr}, and another is to combine the
results of reactor oscillations and the LBL
\cite{Minakata:2002jv}. This section checks possible resolutions
obtainable from the combined analysis of results at telescopes and
results at LBL.

\begin{figure}
{\includegraphics[width=0.75\textwidth]{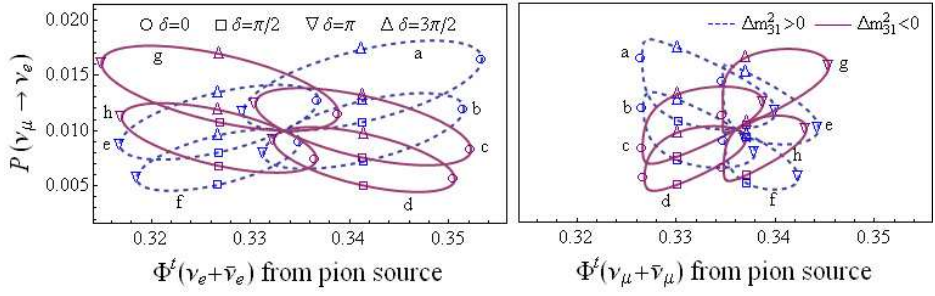}}
\caption{\label{fig:eightpi} Eight loci for $\delta \in [0,2\pi]$
passing a common point in the $\Phi^t(\nu)-P_{\mu e}$ plane.}
\end{figure}

\begin{figure}
{\includegraphics[width=0.75\textwidth]{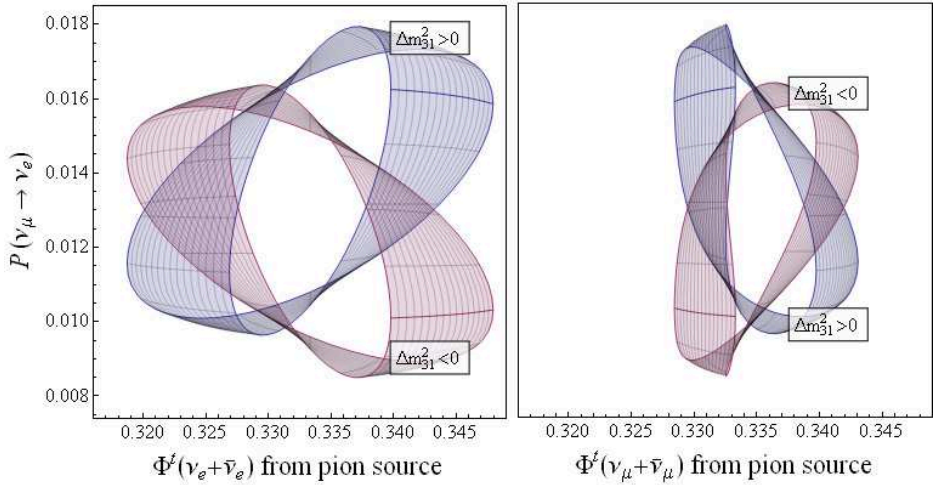}}
\caption{\label{fig:masspi} Dependence of curves on $\Delta
m_{31}^2$: $\theta_{12}$ is allowed within $3\sigma$ CL. }
\end{figure}

The leading term of $P_{\mu e}$ in Eq. (\ref{lblp}) with a
remarkable sensitivity to $\theta_{13}$ does not distinguish
$\Delta m_{31}^2 > 0$ from $\Delta m_{31}^2 < 0$ unless the matter
effect is significant, and any detected flux $\Phi^t(\nu)$ is
fully blind to $\Delta m_{31}^2$. However, the locus of $P_{\mu
e}$ and $\Phi^t(\nu)$, as in FIG. \ref{fig:eightpi} and FIG.
\ref{fig:masspi}, reveals the sign of mass-squared difference
$\Delta m_{31}^2$ due to the phase difference $\Delta
m_{31}^2L/4E_\nu$.

Each single curve in FIG. \ref{fig:eightpi} is obtained by a set
of specific values of $\theta_{23}, \theta_{12}$ and $\Delta
m_{31}^2$, while $\delta$ runs from 0 to $2\pi$. The value of
$\theta_{12}$ is common for all curves in a figure and each of
$\theta_{23}$ and $\Delta m_{31}^2$ has two choices: that
$\theta_{23}$ may be a certain angle or its complementary angle
and that $\Delta m_{31}^2$ may be $|\Delta m_{31}^2|$ or $-|\Delta
m_{31}^2|$. Then, two sets of $(\theta_{13}, \delta)$ exist for
every set of $(\theta_{23}, \Delta m_{31}^2)$, so that total eight
pairs of $(\theta_{13}, \delta)$'s describe the point in $P_{\mu
e}-\Phi^t(\nu_e)$ \cite{GonzalezGarcia:2007ib}. For instance, the
point $(\Phi^t(\nu_e),P_{\mu e})=(0.33, 0.010)$ in FIG.
\ref{fig:eightpi} intersected by the eight curves can be specified
by the following eight pairs of $(\theta_{13}, \delta)$ which
belong to `a - h'curves:
\begin{eqnarray}
    \begin{array}{lll}
        a:~(0.104, 2.3) & ~ & b:~(0.087, 3.8) \nonumber \\
        c:~(0.094, 2.4) & ~ & d:~(0.079, 3.6) \nonumber \\
        e:~(0.086, 0.86) & ~ & f:~(0.071, 5.6) \nonumber \\
        g:~(0.103, 1.01) & ~ & h:~(0.084, 5.4). \nonumber
    \end{array}
\end{eqnarray}
The above values of $(\theta_{13}, \delta)$ are divided into two
groups, according to $\theta_{23}$: [a, b, c, d] for $ \theta_{23} =
\pi/4 -\epsilon$ and [e, f, g, h] for $ \theta_{23} = \pi/4
+\epsilon$ where $\epsilon$ is a positive small angle. They can be
divided also according to the sign of $\Delta m_{31}^2$: [a, b, e,
f] for $\Delta m_{31}^2 > 0$ and [c, d, g, h] for $\Delta m_{31}^2 <
0.$ In other words, the LBL oscillation probability $P_{\mu
e}(\theta_{23}-\pi/4,~ sgn(\Delta m_{31}^2),~ (\theta_{13},
\delta))=0.010$ is degenerated by the following eight possibilities
\begin{eqnarray}
    \begin{array}{llll}
          ~ & P_{\mu e}(-\epsilon,~ +,~ \mathrm{a})
        & = & P_{\mu e}(-\epsilon,~ +,~ \mathrm{b}) \\
          = & P_{\mu e}(-\epsilon,~ -,~ \mathrm{c})
        & = & P_{\mu e}(-\epsilon,~ -,~ \mathrm{d}) \\
          = & P_{\mu e}(+\epsilon,~ +,~ \mathrm{e})
        & = & P_{\mu e}(+\epsilon,~ +,~ \mathrm{f}) \\
          = & P_{\mu e}(+\epsilon,~ -,~ \mathrm{g})
        & = & P_{\mu e}(+\epsilon,~ -,~ \mathrm{h}),
    \end{array}
\end{eqnarray}
where $P_{\mu e}(-\epsilon,~ +,~ \mathrm{a})$, as an example, means
that the probability is obtained when $\theta_{23}-\pi/4
=-\epsilon,~ sgn(\Delta m_{31}^2)$ is $+$, and $(\theta_{13},
\delta)$ is the point on the locus `a'. Likewise, the flux at
telescope $\Phi^t(\nu_e)(\theta_{23}-\pi/4,~ sgn(\Delta m_{31}^2),~
(\theta_{13}, \delta))=0.33$ is degenerated by the following eight
possibilities
\begin{eqnarray}
    \begin{array}{llll}
          ~ & \Phi^t(\nu_e)(-\epsilon,~ +,~ \mathrm{a})
        & = & \Phi^t(\nu_e)(-\epsilon,~ +,~ \mathrm{b}) \\
          = & \Phi^t(\nu_e)(-\epsilon,~ -,~ \mathrm{c})
        & = & \Phi^t(\nu_e)(-\epsilon,~ -,~ \mathrm{d}) \\
          = & \Phi^t(\nu_e)(+\epsilon,~ +,~ \mathrm{e})
        & = & \Phi^t(\nu_e)(+\epsilon,~ +,~ \mathrm{f}) \\
          = & \Phi^t(\nu_e)(+\epsilon,~ -,~ \mathrm{g})
        & = & \Phi^t(\nu_e)(+\epsilon,~ -,~ \mathrm{h}).
    \end{array}
\end{eqnarray}
Thus, a point representing data from two experiments, e.g., the
point $(\Phi^t(\nu_e),P_{\mu e})=(0.33, 0.010)$ or the point
$(\Phi^t(\nu_\mu),P_{\mu e})=(0.34, 0.010)$ in FIG.
\ref{fig:eightpi}, always belongs to many curves specified by
different combinations of the parameters. The number of the loci
passing a point represents the order of degeneracy. In fact, the
order of degeneracy becomes infinite if $\Delta m_{31}^2$ or
$\theta_{23}$ is allowed within a continuous range.

\subsection{Breakable degeneracy by high energy cosmic neutrinos}

FIG. \ref{fig:masspi} displays the loci of $P_{\mu e}$ and
$\Phi^t(\nu)$ for $\theta_{12}$ from 0.51 to 0.69 and $\delta$
from zero to $2\pi$ at $\theta_{13}=0.010$ and
$\theta_{23}=\pi/4$. The projections of the two bands onto
$\Phi^t(\nu)$ axis overlap completely with each other, and even
the projections of the two bands onto $P_{\mu e}$ axis overlap
mostly with each other. So the individual measurement, either
$\Phi^t(\nu)$ or $P_{\mu e}$, is not sensitive to the sign of
$\Delta m_{31}^2$. However, the combination of two measurements as
in FIG. \ref{fig:eightpi} or FIG. \ref{fig:masspi} is
significantly sensitive to whether $\Delta m_{31}^2>0$ or $\Delta
m_{31}^2<0$. On the other hand, if $\theta_{23}$ and $\theta_{13}$
also are allowed within certain ranges rather than with fixed
values, it is hard to get such distinct plots according to the
type of mass hierarchy. Thus, an analysis with a fixed
$\theta_{13}$ as in FIG. \ref{fig:masspi} can be effective only
after a series of reactor or LBL neutrino oscillations, which will
be launched ahead of neutrino detection at telescopes
\cite{Guo:2007ug}-\cite{Itow:2001ee}.

The leading term of $1-P_{\mu\mu}$ in Eq. (\ref{disappear}) is
sensitive to $\theta_{23}$ but does not distinguish
$\theta_{23}=\pi/4-\epsilon$ from $\theta_{23}=\pi/4+\epsilon$. On
the other hand, the flux $\Phi^t(\nu_e)$ in FIG. \ref{fig:shadow},
which is more sensitive to $\theta_{23}$ than the probability
$P_{\mu e}$ is, shows a distinction whether
$\theta_{23}=\pi/4-\epsilon$ or $\theta_{23}=\pi/4+\epsilon$.
Thus, $\Phi^t(\nu_e)$ which is a completely single-valued curve of
$\theta_{23}$ in the currently allowed range as illustrated in
FIG. \ref{fig:twosonu} or in FIG. \ref{fig:shadow} avoids the
degeneracy caused by the complementarity angle of $\theta_{23}$,
unlike $1-P_{\mu\mu}$ in Eq. (\ref{disappear}).

\subsection{Source identification}
\begin{figure}
{\includegraphics[width=0.75\textwidth]{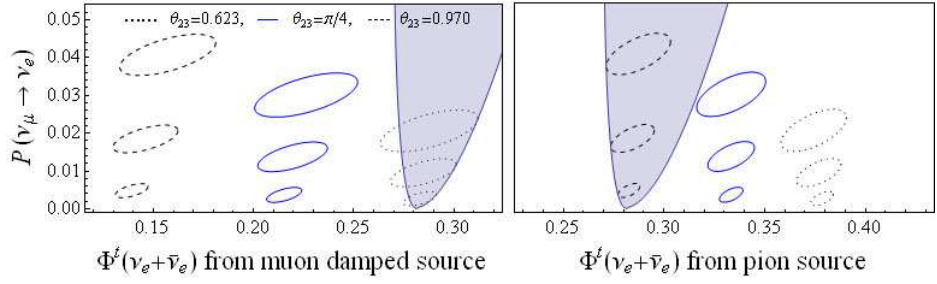}}
\caption{\label{fig:shadow} The loci for $\delta \in [0,~2\pi]$ at
$\theta_{13}=$ 0.15, 0.10, or 0.05 for each $\theta_{23}$.}
\end{figure}
\begin{figure}
{\includegraphics[width=0.75\textwidth]{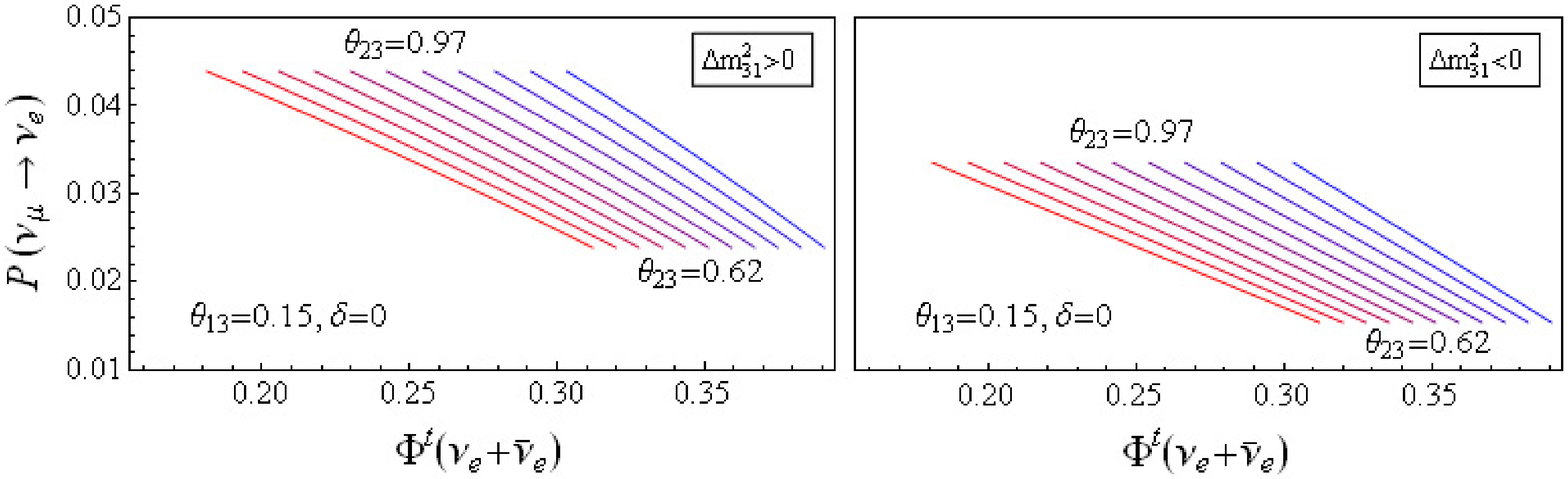}}
\caption{\label{fig:zerodel} The loci for $\theta_{23} \in [0.62,
~0.97]$ for various compositions. }
\end{figure}

\begin{figure}
{\includegraphics[width=0.5\textwidth]{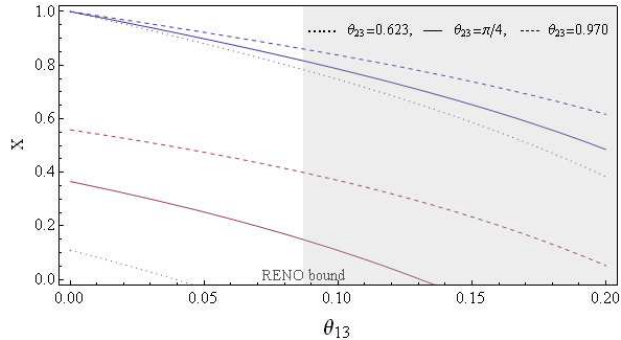}}
\caption{\label{fig:mix} The maximum portion of pion source in a
mixed beam which can be distinguished from the pure pion source. }
\end{figure}

FIG. \ref{fig:shadow} is drawn under the assumption which is that
the neutrino beam was purely from the pion source and purely from
the muon-damped source. In reality, the initial flux in an
astronomical burst is difficult to identify, whether pion is
produced in $p\gamma$ collision or $pp$ collision and whether the
initial condition is pion source, muon-damped source, or a mixture
of them when $p\gamma$ collision is dominant. Unlike the detection
of photons at optical telescopes, the detection of neutrinos at
telescopes should consider the change in neutrino flavors from the
original outbursts. Moreover, broad uncertainties at present in
neutrino mixing angles obstruct telescopes in describing the
initial condition of neutrino beams.

The LBL oscillation probability $P_{\mu e}$ may again share a role
in specifying the source of cosmic neutrinos. The shaded regions
of both panels in FIG. \ref{fig:shadow} show the same area which
both sources cover commonly. So the measurement of $\Phi^t(\nu_e)$
within the shadow may indicate a beam from a pion source (right
panel) for $\theta_{23}=0.970$ or a beam from a muon-damped source
(left panel) for $\theta_{23}=0.623$. The distinction between them
may be obtained by considering $P_{\mu e}$ together with
$\Phi^t(\nu_e)$. For instance, Fig. 5 tells us that $P_{\mu
e}>0.025$ in the shadow can be compatible with $\Phi^t(\nu_e)$
from a pion source but not with $\Phi^t(\nu_e)$ from a muon-damped
source.

FIG. \ref{fig:zerodel} is an example that predicts the relative
composition between a pion source and a muon-damped source, where
$\theta_{12}=0.59, ~\theta_{13}=0.15$, and $\delta=0$. The change
in color of the curves from the blue (rightmost) to the red
(leftmost) indicate the change in composition from 100\% pion
source to 100\% muon-damped source. Unfortunately, the distinction
of the composition by every 10\% as in Fig.\ref{fig:zerodel} lies
far beyond the practicability, since current sensitivities of
detectors are too low to use the strategy in the figure for the
precise comparison.

The detected flux which is partially from pion source and partially
from muon-damped source is expressed in terms of $x$, the portion of
pion-source flux in the total detected $\nu_e$ flux
    \begin{eqnarray}
        \Phi^t(\nu_e) = x \Phi_p(\nu_e) + (1-x) \Phi_{md}(\nu_e),
    \end{eqnarray}
where $\Phi_p$ is a flux from pion source and $\Phi_{md}$ is a flux
from muon-damped source. Even though the exact value of
$\theta_{13}$ was known, the undetermined $\delta$ limits the
composition of mixed flux or the pure pion-source flux to have blind
ranges. FIG. \ref{fig:mix} describes the maximum portion $X$ of
$\Phi_p(\nu_e)$ in a mixed $\Phi^t(\nu_e)$  for various
$\theta_{13}$ and $\theta_{23}$ that can be distinguished from the
pure $\Phi_p(\nu_e)$, where $0<x<X$. The reason of decreasing $X$ as
$\theta_{13}$ increases is because amplitudes of $\delta$ curves
become more sizable. The upper(blue) lines are affected only by the
range in $\delta$, while the lower(red) lines are affected by
systematic uncertainties as well as by $\delta$. The sensitivity of
IceCube detector to astrophysical source was discussed for neutrinos
at TeV to PeV energies \cite{Ahrens:2003ix}. When
$\frac{dN_\nu}{dE_\nu}$ is proportional to $E^{-2}$, the systematic
uncertainty is $+10/-15\%$, and when $\frac{dN_\nu}{dE_\nu}$ is
proportional to $E^{-3}$, the systematic uncertainty is $+5/-20\%$.
In case of $E^{-2}$ spectrum, the composition of a distinguishable
mixture from the pure pion-source beam is drawn by the lower (red)
curves in FIG. \ref{fig:zerodel}. Within the current sensitivity of
the IceCube detector, the mixed flux with more than 40\% (or 20\%)
pion source cannot be distinguished from the pure pion source when
$\theta_{23}$ is 0.970 (or $\pi/4$) and $\theta_{13}$ reaches the
$3\sigma$ upper bound at RENO. Within the current upper bound of
$\theta_{13}$ as seen in the figure, the pure pion source cannot be
distinguished even from the pure muon-damped source, unless $P_{\mu
e}$ does help the identification in such a way as in FIG.
\ref{fig:shadow} and FIG. \ref{fig:zerodel}.

\section{\label{sec:sec5}Concluding remarks}

Neutrino telescopes will detect neutrino beams out of astronomical
bursts. However, the results to obtain are barely helpful in
describing the initial condition of cosmic neutrino beams since
the uncertainties in neutrino masses and mixing angles are broad
and the fluxes to be measured are sensitive to the masses and the
mixing angles. On the other hand, the improvement of the precision
in parameters is also hard to attain by using results at neutrino
telescope itself since even an original beam as initial condition
cannot be defined without using the telescope.

We took a strategy to consider the fluxes to be detected at a
telescope like IceCube in company with oscillation probabilities at
a LBL T2K. The expected fluxes are examined for the sensitivities to
mixing angles, in comparison with the sensitivities of other types
of oscillation probabilities to mixing angles. A few restricted
cases were presented as examples to show that neutrino fluxes at
telescopes may be useful to resolve the degeneracies embedded in
terrestrial neutrino oscillations. It was followed by the discussion
on the limit of source identification which is allowed within the
sensitivity of IceCube to astrophysical neutrinos.

\begin{acknowledgments}
K. Siyeon thanks Z. Xing for information on neutrino telescope and
thanks physicists at IHEP in Beijing for warm hospitality. This work
was supported by the Korea Research Foundation Grant funded by the
Korean Government(MOEHRD).(KRF-2005-041-C00108)

\end{acknowledgments}


\end{document}